\documentclass[reprint,
superscriptaddress,
amsmath,amssymb,aps,showkeys,showpacs,
twoside,final,secnumarabic,
nofootinbib]{revtex4-2}

\usepackage[paperwidth=205mm,paperheight=290mm,top=17mm,bottom=25mm,
inner=17mm,outer=17mm,
twoside]{geometry}

\usepackage{cmap} 
\usepackage[T1,T2A]{fontenc}
\usepackage[utf8]{inputenc}
\usepackage[russian,english]{babel}
\usepackage{color}
\usepackage{graphicx}
\usepackage{dcolumn}
\usepackage{bm} 
\usepackage[unicode=true,colorlinks=true,linkcolor=magenta, urlcolor=blue, citecolor = blue,breaklinks]{hyperref}
\usepackage{multirow}
\usepackage{url}
\usepackage{breakurl}
\DeclareGraphicsExtensions{.eps}

\newcount\issue
\newcount\Vol
\newcount\numb
\usepackage{fancyhdr} 
\def\Vol{\textbf{80}}
\def\numb{x}
\newcommand{\cD}{\ensuremath{\mathcal{D}}}

\usepackage{xcolor}

\usepackage{lineno}

\begin{document}

\title{
ELVESLOCATION:\\ A NEW APPROACH FOR STUDYING THUNDERSTORM ATMOSPHERE
} 

\def\addressa{Faculty of Physics, Lomonosov Moscow State University, Moscow, 119991, Russia}
\def\addressb{Skobeltsyn Institute of Nuclear Physics, Lomonosov Moscow State University, Moscow, 119991, Russia}

\author{\firstname{S.A.}~\surname{Sharakin}}
\email[E-mail: ]{sharakin@mail.ru}
\affiliation{\addressb}
\author{\firstname{R.E.}~\surname{Saraev}}
\email[E-mail: ]{saraevrom@gmail.com }
\affiliation{\addressa}
\affiliation{\addressb}

\begin{abstract}
Thunderstorm electrical discharges, which vary in their type of initiation and development (including compact intracloud discharges and initial breakdown pulses), also manifest differently as recorded electromagnetic signals in both the radio and optical ranges. A unique ionospheric optical imprint of a discharge is an ELVES – a sub-millisecond glow at altitudes around 90 km with a characteristic spatiotemporal pattern in the form of an expanding ring, reaching a diameter of several hundred or even a thousand kilometers. The position of the elve, its expansion velocity, the azimuthal and radial distribution of the glow intensity (and their temporal dynamics) contain information about the location and orientation of the discharge, as well as its current function. This information is most fully represented in the data recorded by orbital detectors of dynamic images with a wide field of view and microsecond temporal resolution, such as the TUS and Mini-EUSO. Recovering the parameters of a discharge from orbital detector data is associated with solving a complex inverse problem, since the phenomenon of interest is mediated by several processes, including the interaction of the discharge's electromagnetic pulse with the ionosphere and the non-trivial instrument response function of the detector. These additional sources of uncertainty are most consistently accounted for within the Bayesian paradigm. The application of simulation-based inference (SBI) allows for the reconstruction of discharge parameters within a dynamic model, thereby taking a step towards creating a new method for studying the thunderstorm atmosphere -- elveslocation. This work presents a dynamic model of an elve, which enables both the identification of characteristic spatiotemporal patterns of an event (generator mode) and the reconstruction of discharge parameters (SBI simulator mode).

\end{abstract}

\pacs{Suggested PACS}\par
\keywords{lightning discharges, transient luminous events, ELVES, orbital detectors, elves-location, Bayesian inference.\\[5pt]}

\maketitle
\thispagestyle{fancy}

\section{Introduction}

Research on thunderstorm atmospheres and related phenomena is a crucial source of information about a wide variety of physical processes, mainly those associated with lightning~\cite{Dwyer2014}. The properties of a long spark obtained under laboratory conditions cannot always be directly transferred to its natural analog—the lightning discharge. The processes of both the development of the lightning channel itself and its initiation may differ. Therefore, enormous attention continues to be paid to various, sometimes quite exotic, methods of ``extracting'' information about thunderstorm processes. For instance, to observe the development of a leader channel at a given moment in time and in a given area of space, it has even been proposed to ``bombard'' rain clouds with beams of high-energy protons~\cite{Gorev2024}.

Traditional methods for studying discharge processes in thunderclouds include optical recording of the glow from a lightning channel (leader or return stroke) emerging from the cloud using special photo and video cameras, and numerous methods for detecting the discharge's radio signal, primarily magnetic field direction finding and the time-of-arrival method~\cite{Rakov2003}. If the goal is to image the developing channels of any type of lightning discharge, methods such as the time-of-arrival in the VHF/UHF ranges and VHF/UHF interferometry can be used~\cite{Rison2016}.

One of the most enigmatic and unresolved questions in thunderstorm physics is elucidating the mechanisms of lightning initiation~\cite{Dwyer2014}. Various options are proposed, see, for example, \cite{Petersen2008}, \cite{Kostinskiy2020}, \cite{Sysoev2025}, and it is gradually becoming clear that such a mechanism must include a detailed description of the many different discharge processes observed in the thunderstorm atmosphere. In~\cite{Kostinskiy2020}, among the facts requiring explanation, particular attention is paid to identifying several stages of lightning preceding the currently well-understood stepped leader stage, and to the radio and optical characteristics of various kinds of discharge recorded during these preliminary stages.

The onset of the stages preceding lightning is associated with an initiating event (IE). One type of~IE is known as \textit{compact intracloud discharges} (CIDs), which were discovered back in the 1980s by measuring the radiation component of the electric field in broadband radio signals~\cite{LeVine1980}. Due to their characteristic waveform and short field pulse duration (not exceeding several tens of microseconds~\cite{Karunarathne2015}), they are also called narrow bipolar events (NBEs).

CID/NBE is the most powerful natural source in the~VHF range: its power in the 60-66~MHz band can reach 300~kW. The currents of these discharges are estimated to be tens of kiloamperes, with recorded peak values even exceeding 100~kA, which is typical only for very strong return strokes. To date, it remains unclear by what mechanism such a short channel (less than 1~km) can concentrate such a large charge (estimated at up to 0.5-1~C~\cite{Rison2016}) in such a short time. Another point requiring explanation is the absence of optical emission from CIDs.

Studies have been conducted to investigate the context of~CIDs, estimating the proportions of events isolated from the return stroke and those preceding it~\cite{Leal2019}. Some isolated CIDs are considered precursors to cloud-to-ground lightning, since lightning discharges were recorded in nearly the same location after intervals of seconds. According to refined data, it has been established that the majority of intracloud and negative cloud-to-ground lightning is not initiated by~CIDs, but by much less powerful events~\cite{Lyu2019}, \cite{Bandara2019}.

A separate mystery is associated with the determination of the altitude of CIDs. The altitude distribution of CIDs measured by the Los Alamos Sferics Array (LASA) turned out to be very broad, covering altitudes from~2 to 30~km~\cite{Smith2004}. Given that even the most powerful thunderclouds do not develop above 30~km, it is difficult to explain the occurrence of CIDs there. However, the accuracy of these CID altitude estimates has been questioned by several authors. In particular, subsequent studies (with more precise localization methods, albeit with smaller event statistics) indicate a narrower altitude distribution for CIDs, with an upper limit of~18–20~km~\cite{Leal2019}, \cite{Karunarathna2015}.

In addition to~NBEs, series of discharges known as \textit{initial breakdown pulses} (IBPs) are often recorded just before the return stroke. These are characterized by much stronger and longer-lasting VHF signals than the stepped leader~\cite{Rhodes1989}. IBPs are series of bipolar microsecond-scale electric field pulses with amplitudes comparable to or even exceeding that of the return stroke. The most pronounced, so-called ``classic'' IBPs, have range-normalized (to 100~km) field amplitudes averaging about 1.5~V/m but can reach up to~8.5~V/m~\cite{Smith2018}. Measurements using high-speed video cameras of bright flashes synchronized with IBPs~\cite{Stolzenburg2013} indicate that this type of discharge is produced by a hot, highly conductive and intensely luminous plasma that has accumulated a significant charge. The Event-by-event analysis of the IBP fields and the estimation of the parameters of the generating discharge are presented in~\cite{Karunarathne2014}, \cite{Karunarathne2020}.

In modern theoretical models of lightning initiation, both CIDs and IBPs play a crucial role. The development of such models is largely based on the quantitative estimation of various discharge parameters and their interdependence. A potential challenge is the insufficient accuracy and reliability of existing estimates, and addressing this may involve, among other things, the development of new methods of location intracloud discharges. Perhaps even the very concept of their ``location'' could have an alternative meaning. The waveform of a single IBP pulse exhibits extremely strong jaggedness, which may indicate that individual peaks arise from the interaction of plasma structures (leaders or networks) rather than from the radiation of a single plasma channel~\cite{Kostinskiy2020}.
A new fractal mechanism for the generation of CIDs is presented in~\cite{Iudin2018}, describing it as cluster-cluster aggregation of dynamic branching discharge structures within a thundercloud.

The interpretation of ground-based observations obtained via the VHF/UHF methods is not always unambiguous. A highly promising approach involves the use of satellite-based optical and radio sensors and the correlating of their signals with data from ground-based stations. For example, CIDs were detected via their VHF signal by the ALEXIS satellite as paired transionospheric pulses caused by reflections from the Earth's surface~\cite{Holden1995}. A comprehensive study of CIDs/NBEs was conducted using data from the FORTE satellite combined with ground-based NLDN and LASA networks~\cite{Jacobson2012}.

In addition to radio signals, lightning discharges create a series of rapidly changing optical flashes in the upper atmosphere, referred to in the literature as transient luminous events (TLEs). Initially, optical emissions were recorded from large, tens of kilometers in diameter volumetric electrical atmospheric discharges at altitudes of 50-70~km~\cite{Franz1990}; this type of TLE was later named ``sprites''.

Shortly thereafter, brief flashes of upper atmospheric glow at altitudes of 85-90~km, coinciding in time with lightning strikes, were recorded from the space shuttle Discovery~\cite{Boeck1992}. The region of enhanced luminosity had a thickness of~10 to 20~km and a diameter of about 500~km, and its brightness was approximately twice that of the background atmospheric glow. Later, these submillisecond glows were measured in detail during the SPRITES'1995 campaign~\cite{Fukunishi1996} and were named ELVES (Emission of Light and Very low frequency perturbations due to Electromagnetic pulse Sources). This type of TLE is often referred to simply as ``elve'' (the singular form of ``elves'').

Indeed, TLEs represent an additional channel for extracting information about a lightning discharge. The luminosity of an elve is sensitive to the magnitude of charge transferred by the discharge~\cite{Cummer2005}, while the analysis of an elve's development can provide detailed information about the discharge's geometric characteristics - its localization and orientation.

The mechanism for elve formation was first proposed in~\cite{Inan1991} to explain the early, or fast, subionospheric VLF perturbations. Since then, a large number of studies (a brief review is provided in the next section) have been published, dedicated to models of TLE generation (forward models).

In this work, we demonstrate that the application of Bayesian methods effectively solves the inverse problem: estimating the parameters of a lightning discharge (including intracloud ones)—primarily its location and orientation—from its ionospheric optical imprint. The high-precision information obtained in this way can be highly valuable for constructing discharge models and refining the mechanism of lightning initiation. The most detailed data on elves are provided by satellite-based wide-angle imaging detectors with high (microsecond) temporal resolution. Our work specifically shows how the characteristic features of the instrumental function of such detectors can be taken into account during elve reconstruction within the framework of simulation-based probabilistic inference~(SBI).

\section{RESEARCH ON ELVES}

The first detailed measurements of elves from low Earth orbit were performed by the ISUAL photometer system~\cite{Chern2003}, \cite{Chen2008}. Observations of ELVES and other TLE types from orbit were also conducted as part of the GLIMS~\cite{Sato2015} and ASIM~\cite{Neubert2019} projects. \cite{Neubert2021} presents interesting results from ASIM spectrophotometric observations of the upper parts of thunderclouds. In addition to short ($\sim10$~$\mu$s) light pulses at 337~nm, which the authors identify as CIDs, discharges of a special type—blue jets—directed upward from the cloud top, as well as elves, were recorded. Unfortunately, the two cameras in ASIM's scientific payload, while having good spatial resolution ($10^6$ pixels with a side length of 400~m at sea level), record signals with a temporal resolution of $\sim83$~ms, which is too low for studying elve dynamics.

Theoretical models of the 1990s-2000s (see, for example, \cite{Inan1996}, \cite{Barrington-Leigh2001}) showed that an elve can be represented as the result of a chain of events: a lightning discharge generates an electromagnetic pulse (EMP), which propagates and causes ``heating'' of electrons, ionization, and excitation of air molecules (primarily nitrogen) with subsequent light emission. The characteristic spatiotemporal pattern of the ionospheric glow region—a ring expanding at superluminal speed—is formed as a result of the intersection of the approximately spherical EMP wavefront with the ``plane'' of the elve, located at an altitude of about 90~km. (Calculations confirmed by experiments showed that the thickness of the glow region is much smaller than its transverse spatial dimensions.)

This glow pattern contains detailed information about the discharge: a) the center of the ring and its (variable) expansion velocity – about the location of the discharge; b) the azimuthal (along the ring) distribution of glow intensity and its change over time – about the orientation of the discharge; c) the radial distribution of the glow and its variations over time – about the current function, discharge multiplicity, etc.

We note that temporal changes are present in all the listed characteristics. That is, the most significant source of information is not separately taken spatial or temporal data, but precisely spatiotemporal patterns. Therefore, suppliers of such data can be multi-channel photometers with high temporal resolution – dynamic imaging detectors. However, for the data recorded by these detectors to be truly informative, two main conditions must be met.

1. The detector must have a wide field of view (FoV) and a sufficiently high spatial (angular) resolution to recognize geometric and photometric patterns in the signal.

2. The detector must record the signal (image) with high temporal resolution (and, consequently, possess high sensitivity), sufficient to reveal temporal dynamics.

The recorded data contain information about the discharge if the detector's optics form an image of a significant portion of the elve's ring—in the atmosphere at an altitude of 90~km, this corresponds to scales ranging from several tens to hundreds of kilometers. The required temporal resolution of the instrument is determined by both the characteristic timescales of the discharge development and the apparent expansion speed of the ring and in typical situations amounts to units of microseconds.

Precisely this type of instrument has been developed over the past two decades by the JEM-EUSO\footnote{Joint Exploratory Missions for an Extreme Universe Space Observatory.} collaboration~\cite{Casolino2017} for the orbital detection of ultra-high-energy cosmic rays (UHECRs) \cite{Coleman2023}. The extreme rarity of UHECRs and the need to detect faint sources moving at the speed of light led to the development of optics and a photodetector meeting the above requirements—a FoV of tens of degrees and microsecond resolution.

Multi-channel recording of dynamic elve images with high temporal resolution (0.8~$\mu$s) from orbit was first achieved by the TUS detector, deployed on the Lomonosov satellite in 2016-2017~\cite{Klimov2019}. In particular, double elves (duplets) were detected~\cite{Kaznacheeva2019}, where the second ring is associated with the impact on the ionosphere of an electromagnetic pulse (EMP) reflected from the conductive surface of the Earth.

A veritable laboratory for studying elves has become the Mini-EUSO orbital detector – a wide-angle telescope operating aboard the International Space Station since~2019~\cite{Bacholle2021}. Mounted inside the~ISS next to a UV-transparent window, this telescope observes the Earth's atmosphere within a wide FoV ($\sim40^\circ$ in diameter with a nadir-pointing axis) with a temporal resolution of $\tau=2.5$~$\mu$s. The projection of its FoV at the typical elve emission altitude covers an area of about $250\times250$~km, whereas the FoV of a single detection channel at this altitude represents a ``pixel'' of $\sim4.7\times4.7$~km. High temporal resolution and a large number of UV-sensitive channels (2304) allow for detailed tracking of an elve's dynamics at various stages of its development.

\begin{figure*}
    \centering
    \includegraphics[width=0.45\textwidth]{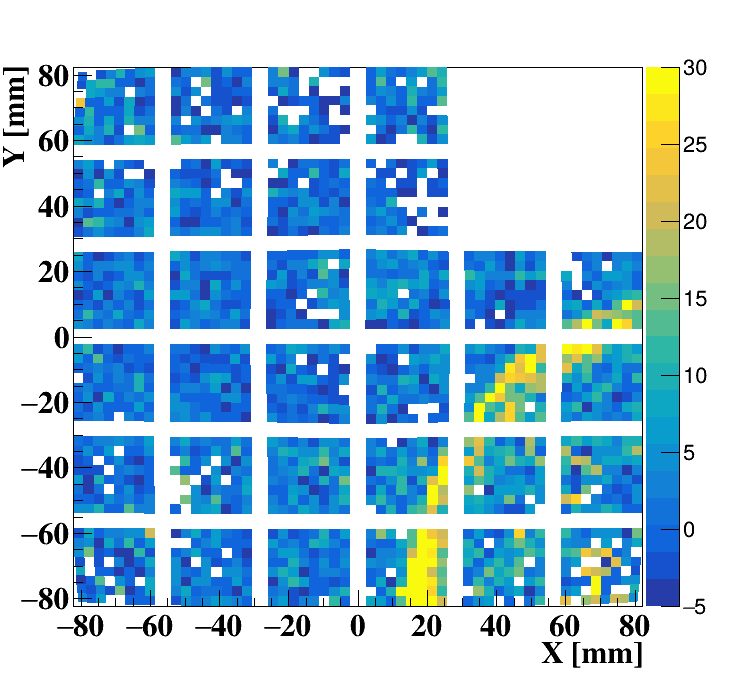}
    \includegraphics[width=0.45\textwidth]{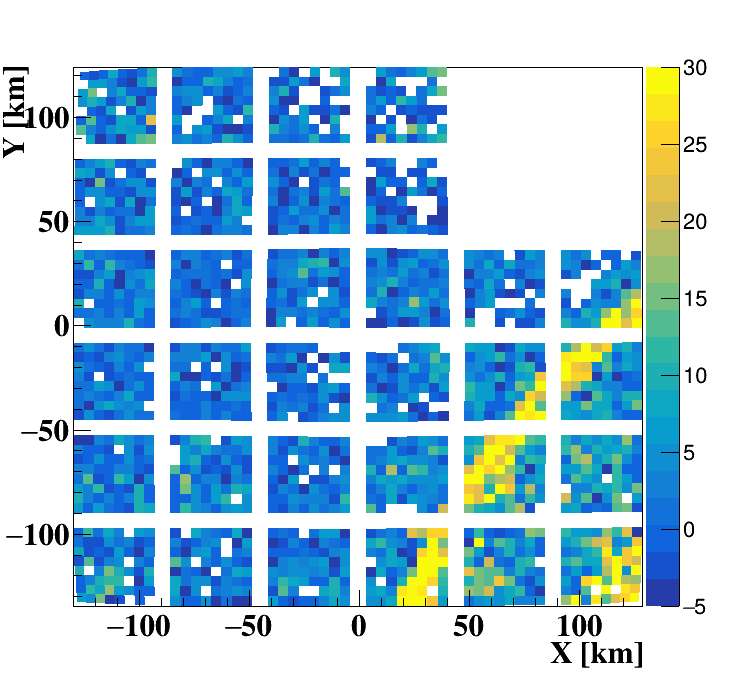}
    \caption{Distortion of the Mini-EUSO image due to variations in time delays across the FoV. Left — image of one frame of the ELVES20191205 event on the photodetector, right — its projection onto an altitude of 90~km.}
    \label{fig:MEE_Event2}
\end{figure*}

If the discharge that generated the elve is located near the instrument's FoV, the dynamic image recorded by Mini-EUSO corresponds most closely to the actual spatiotemporal excitation pattern of the ionosphere.
For distant discharges, when the development of the elve is observed in its later stages, the transformation of the glow pattern at the edge of the detector's FoV can be significant. In Fig.~\ref{fig:MEE_Event2}, the same event is presented in two formats: on the left - as an image of a single frame on the photodetector, on the right - after its projection onto a spherical surface of radius\footnote{With its center at the Earth's center, here $R_\mathrm{E} = 6371$~km is the Earth's mean radius.} $r = R_\mathrm{E} + H_\mathrm{e}$, where $H_\mathrm{e} = 90$~km. In the latter case, it is taken into account that the time delay for each direction within the FoV (in each pixel) differs; the maximum difference — between pixels in the corner and the center of the photodetector — is about 150~$\mu$s. This variability in time delays leads to a noticeable ``deformation'' of the spatial pattern as well; in particular, the curvature of the elve's arc changes significantly.

In addition to this purely kinematic transformation, distortions introduced by the detector's instrumental function also play a significant role. The figure clearly shows numerous dead zones in the detector's FoV, associated with structural gaps between the blocks of photoregistering sensors—multi-anode PhotoMultiplier Tubes (PMTs). An individual PMT contains an array $8\times8$ of sensor channels, each with a square input window on the $\sim3$~mm side. The gap between PMT is 4-5~mm. Another feature of the Mini-EUSO photodetector is its automatic switching of $2\times2$ PMT blocks to a reduced sensitivity mode in case of strong illumination (in Fig.~\ref{fig:MEE_Event2}, the upper right block appears ``switched off'').

Despite these noted characteristics of the dynamic image formed by the orbital detector, it contains all the types of characteristic information listed above. This allows for estimating important parameters of both the elve itself and the discharge that generated it. Essentially, this points to the prerequisites for creating a new method of discharge reconstruction based on its ionospheric ``imprints''. We will call this reconstruction method \textit{elveslocation}.

\section{BAYESIAN RECONSTRUCTION}

The high-precision reconstruction of discharge parameters requires not only instruments with high spatiotemporal resolution but also appropriate methods to analyze the data they provide.

Simple fitting (parameter optimization) of the data recorded by the detector often relies on strong simplifications of the interpretational model, various forms of data factorization during pre-processing (e.g., extracting geometric or kinematic data blocks), semi-empirical selection of summary statistics, etc. In more complex models involving a large number of auxiliary parameters, fitting is performed by substituting various point estimates for these additional parameters (plug-in solutions using expert estimates, maximum likelihood estimates, etc.). This leads to an unjustified narrowing of the uncertainty intervals of the obtained estimates and sometimes to their bias, i.e., to an overestimation of the actual accuracy and reliability of the reconstruction method.

It is important to emphasize that the reconstruction problem considered here is an inverse problem. In it, the phenomenon of interest, an electrical discharge in a thundercloud, is connected to the observed data through a whole chain of intermediate processes.

Beyond the discharge itself and the generation of an electromagnetic pulse (EMP), these include:
\begin{itemize}
\item The interaction of the EMP with the ionosphere (``heating'' of electrons, changes in their concentration due to ionization and dissociative attachment).
\item The excitation of nitrogen and oxygen followed by emission in characteristic lines and bands (primarily 1P and 2P of~N$_2$, 1N and 2N of N$_2^+$, 1P of O$_2$).
\item The propagation of radiation in the atmosphere.
\item Transformation of the signal in the detector (instrumental response function).
\item Potential signal preprocessing steps (flat fielding, approximations of active signal and background, background subtraction, etc.).
\end{itemize}
At each of these intermediate stages, new sources of uncertainty are introduced into the model. This makes solving the reconstruction problem by ``simple inversion'' impossible and necessitates the use of adequate quantitative methods for uncertainty assessment.

The most consistent way to account for these uncertainties, in our view, is probabilistic modeling. In this framework, all parameters of the model (both primary and auxiliary) become stochastic variables characterized by their own probability distributions. In this case, the experiment serves as a source of data $\cD_\mathrm{obs}$ that refines our information about these parameters~$\Theta$. The reconstruction problem is then formulated as probabilistic inference (Bayesian inference) based on Bayes' theorem:
\begin{equation}
\label{eq:Bayes}
p(\Theta|\cD_\mathrm{obs}) = p(\cD_\mathrm{obs}|\Theta) p(\Theta) / p(\cD_\mathrm{obs})					
\end{equation}
Here, $p(\Theta)$ and $p(\Theta|\cD_\mathrm{obs})$ represent the information about the model parameters available to us before conducting the experiment (a priori) and after it (a posteriori), expressed in the form of the corresponding probability distributions, the prior and posterior distributions.

More precisely, in the presence of auxiliary parameters~$\eta$, marginalization must be additionally performed:
\begin{equation}
\label{eq:Margin}
p(\Theta|\cD_\mathrm{obs}) \propto \int d\eta\,  p(\cD_\mathrm{obs}|\Theta,\eta)\,  p(\Theta,\eta)			\end{equation}
i.e., averaging (rather than optimization or substitution of some other suitable point estimate) over these additional sources of uncertainty.

The factor $p(\mathcal{D}_\mathrm{obs} | \Theta)$ or $p(\mathcal{D}_\mathrm{obs} | \Theta, \eta)$ appearing in~(1) and~(2) is called the likelihood function (as a function of the parameters for given observations), marginal, and complete, respectively. It is precisely this function that expresses the connection between the model parameters and the measured data. In ``classical'' reconstruction approaches, it also arises naturally – as a function that needs to be maximized, thereby finding the parameter values (MLE, maximum likelihood estimation).

Supplementing the likelihood function with a prior distribution allows a transition from point estimates to a full-fledged probability measure in the parameter space. This transition not only regularizes the search for the extremum (the mode of the posterior distribution – the so-called MAP estimate) but also enables accounting for the numerous uncertainty sources of the extended model and formulating reconstruction results in the form of probability intervals (most commonly, the so-called HDI, highest density interval)~\cite{Loredo2024}.

Bayesian data analysis began to be actively applied in science from the 1980s onward (see, for example,~\cite{Sivia2006}). However, calculating posterior distributions according to equations~(1) and~(2) was possible only for a very limited class of relatively simple models. A significant impetus for the widespread use of Bayesian methods in solving scientific and applied problems came after the development of user-friendly (``high-level'') libraries that implement computations using Markov Chain Monte Carlo (MCMC) sampling methods, such as STAN, PyMC, TensorFlow Probability\footnote{https://www.pymc.io, https://mc-stan.org, https://www.tensorflow.org/probability}, and others. These libraries allow one to formulate a probabilistic model (i.e., specify prior distributions and likelihood functions) in a high-level language, and the computation of samples from the posterior distribution (``samples'') occurs automatically (with its own set of parameters controlling the sampler algorithm, of course). Due to their proximity to high-level programming languages, such approaches have been termed probabilistic programming.

In the case of elveslocation, the parameters~$\Theta$ may include:
the discharge location coordinates\footnote{In a Cartesian coordinate system, for example, one tied to the detector's nadir point for orbital detection.} $x_0$, $y_0$, $z_0$;
 polar $\alpha_0$ and azimuthal $\phi_0$ angles of its dipole moment orientation;
characteristic rise and fall times, and other parameters of the current function.

The set of auxiliary parameters~$\eta$ depends on the choice of the interpretational model (i.e., the framework within which the reconstruction is performed) and can be quite extensive. It may include
characteristics of the ionospheric layer near the elve generation altitudes (e.g., the electron concentration profile), features of light scattering during its propagation to the detector, and, of course, the instrumental response function, modeling the detector's response. The latter must account for both the focusing properties of the optics (the point spread function, PSF) and the characteristic sensitivity distribution of the photodetector across the entire FoV (sensor sensitivities, the presence of ``dead'' zones, etc.).

We previously proposed a simple kinematic model of an elve to reconstruct Mini-EUSO elves~\cite{Sharakin2024-KI}. Its primary goal was to demonstrate how, within a Bayesian framework, one can reconstruct an elve without  factoring out geometric data (i.e., analyzing instantaneous images). During the preprocessing stage, for each ELVES event, a dataset was formed consisting of the signal peak time in each detector channel: $\cD_1 = \{ T_p[i] \}$, where $i$ is the channel identifier.

To probabilistically describe these data, it is sufficient to introduce a set of elve parameters into the model, $\Theta_\mathrm{kin} = \{ x_0, y_0, z_0 \}$, and complement them with auxiliary parameters $\eta = \{ T_0, \sigma_\mathrm{mes} \}$, characterizing time synchronization and measurement time error. This model assumes that the elve emits light at a fixed altitude $H_\mathrm{e}$, i.e., it neglects the thickness of the emitting layer (estimated at 5-10~km according to~\cite{Inan1996}).

Furthermore, the model relies on a significant assumption: the signal peak time in a channel corresponds to the moment the EMP wavefront passes through the center of that channel's FoV. This hypothesis was justified by the approximate symmetry of the PSF and the negligible change in the signal during the wavefront's traversal of the channel's FoV. Of course, both assumptions require appropriate verification and, if necessary, refinement (modification) of the model, which is fully in line with the spirit of the Bayesian approach (for more on the Bayesian workflow, see~\cite{Gelman2020}). In particular, one promising modification of the model could be the introduction of a random offset of the peak from the center of the FoV (e.g., with a common parameter for all channels describing the width of the offset distribution).

Despite the obvious simplicity of this model, it allowed one not only to estimate the discharge location, but also to draw some conclusions about its plausibility. In particular, using~$H_\mathrm{e}$ (emission altitude) as one of the parameters and comparing the model predictions with different prior uncertainties for $H_\mathrm{e}$ made it possible to assess the informativeness of certain events not only with respect to the discharge altitude~$H_0$ (which, due to the sphericity of the Earth, generally differs from the Cartesian coordinate~$z_0$) but also with respect to~$H_\mathrm{e}$.
Unfortunately, in most cases, the presence of a strong linear correlation in the posterior distribution $p(H_0, H_\mathrm{e} | \cD_1)$ imposed significant limitations on reconstruction accuracy: the likelihood function of the kinematic model is informative only with respect to the combination $h_\mathrm{e} = H_\mathrm{e}-H_0$. This is in part due to the insufficiently complete extraction of information from the recorded data, indicating the need to apply more complex models~\cite{Sharakin2024}.

\section{DYNAMIC RECONSTRUCTOR MODEL}

Data from orbital detectors such as TUS or Mini-EUSO consist of digitized signal values~$S$ in each channel of the photodetector (after sensitivity normalization and background subtraction), $\cD_2 = \{ S_k[i] \}$. Here, as before, $i$ identifies the channel (and thus the portion of the FoV it ``observes''), while the index~$k$ numbers the sampling time steps, $k=1,\ldots, K$. The total duration $K$ can be determined by the memory allocated for the event ($K = 256$ for TUS events) or fixed during the pattern recognition stage (identifying active channels and the corresponding time window).

As noted above, the data~$\cD_2$ undoubtedly contain information not only about the location of the discharge that generated the elve but also about its orientation and current function. To extract this information from~$\cD_2$, an interpretational model is required, one that accounts for the specific features of the interaction of the EMP with the ionosphere and the signal formation in the detector.

Forward models of lightning EMP interaction with the ionosphere have become a primary focus of many studies, see, for example,~\cite{Rowland1996} (a 2.5D model with separate simulation of vertical and horizontal discharges) and~\cite{Cho2001} (a 3D model of intracloud discharges), and references therein. The work~\cite{Kuo2007} is dedicated to modeling the response of the orbital ISUAL detector to an elve generated by lightning.

We would like to separately highlight the series of works by the Stanford group led by Umran Inan, who have been consistently developing this field for many years:
\cite{Taranenko1993a},\cite{Taranenko1993b} – a 1D model of ionization and glow within a kinetic framework;
\cite{Inan1996} – a 2D-cylindrical model for a vertical discharge;
\cite{Veronis1999} – an extended 2D model with quasi-electrostatic field;
\cite{Marshall2010} – a 3D model that incorporates the geomagnetic field.
A quantitative analysis of model data with typical ``input'' characteristics of intracloud and cloud-to-ground discharges in~\cite{Marshall2010} led to the following conclusions:\\
1) Discharge parameters (altitude, amplitude, duration, and orientation), as well as electron concentration and geomagnetic field, significantly influence the interaction of an EMP with the ionosphere and these effects are interdependent.\\
2) The sequence of discharges characteristic of most intracloud lightning can cause a cumulative disturbance in electron concentration. Each of these discharge groups creates a potentially observable optical event in the ionosphere—a sequence of ``flashing elves''—detectable by ground-based or satellite photometers.

This series of works was further extended by incorporating the detector component into the model in~\cite{Marshall2012} and~\cite{Marshall2015}. In these studies, Robert Marshall compares the results of detailed modeling within a~3D framework with real data recorded by the ground-based detector PIPER.
During the summer 2008 observational campaign at the Langmuir Laboratory, the PIPER detector measured 803 elves, 78\% of which were associated with negative cloud-to-ground discharges~\cite{Newsome2010}. The detector's photodetector consisted of only two photometer modules – a ``horizontal'' and a ``vertical'' line, each with 16 channels (compare this with~256 channels in TUS and~2304 in Mini-EUSO).
In addition to qualitative comparison of model and real data (and fitting discharge parameters), an attempt was made to search for summary statistics of the data that could serve as a basis for reliable reconstruction in the future. Essentially, these works represented an attempt to develop a method for elveslocation using data from ground-based detectors within a non-Bayesian approach.

Unfortunately, the insufficiently high temporal resolution (signal integration time of 40~$\mu$s) and observation conditions that are not optimal (from the perspective of ``decoding'' the data) significantly complicate the application of Bayesian methods to such data.
Due to the ``superluminal expansion'' of the elve's ring, in a ground-based detector observing from below and to the side, the event is presented in a ``heavily encoded'' form: the arrival times of photons from different parts of the elve are highly entangled (in~\cite{Inan1996}, the terms ``temporal focusing and defocusing'' are even used for this).
In contrast, orbital detectors oriented towards the nadir only weakly ``deform'' the spatiotemporal pattern of the ionospheric imprint and allow the use of relatively simple probabilistic models to extract information about the discharge. An example of such a dynamic model is presented below.

When formulating the interpretational model of an elve used to reconstruct discharge parameters from data~$\cD_2$, we initially based our approach on a set of assumptions that significantly simplify the model of the EMP-ionosphere interaction compared to the models in~\cite{Marshall2010}; \cite{Marshall2012}. Our primary interest was accounting for the changes introduced into the signal by the detector's instrumental response function.
As in the kinematic model, we assumed that the thickness of the layer where the primary nitrogen excitation and emission occur can be neglected. Experiments and detailed modeling estimate this thickness to be about 5~km (compared to~$H_\mathrm{e}\sim90$~km), which is roughly the size of the FoV of a Mini-EUSO channel at that altitude. This model inaccuracy (model misspecification) can be interpreted as a modification of the current function (i.e., the dependence of the discharge channel current on time, $I(t)$).
Consequently, for the current function we use only fairly general, ``coarse'' profiles consisting of two edges: a current rise and a current decay, with characteristic times $\tau_\mathrm{r}$ and~$\tau_\mathrm{d}$. The shape of these profile segments can be chosen from several classes: linear (LIN), exponential (EXP), Gaussian (GAUS), which aligns well with previous work (e.g.,~\cite{Marshall2012} uses a LIN+GAUS profile).
Another useful option for the current function in analysis is the ``instantaneous discharge'', where the duration of the EMP front can be neglected compared to other characteristic times in the model.

Another assumption concerned the process of optical emission generation. Here, we used the simplest variant of a proportional ionospheric response: each unit ``patch'' of the elve at altitude~$H_\mathrm{e}$ emits isotropically with a power proportional to the power of the incident EMP (which, in turn, is proportional to $|dI/dt|^2$ in the far field). Although the EMP interaction with the ionosphere is nonlinear (including due to the presence of ionization thresholds), estimates are provided in~\cite{Marshall2010}; \cite{Marshall2012} allow us to consider these deviations from linearity as small; they can also be ``hidden'' within the model misspecification of the crude approximation of~$I(t)$.

The portion of radiation from the elve patch that enters the telescope's entrance pupil (neglecting scattering in the atmosphere at altitudes above 90~km) was distributed among the individual detector channels, taking into account the FoV of each. To account for the specifics of signal detection, the detector's instrumental response function~$\varepsilon_\mathrm{d}$ was introduced. In general, this is a function of the field angles $(\gamma,\psi)$ pointing from the detector~D to the observed ``point''~E on the~elve. This function incorporates both the contribution of the Point Spread Function (PSF) and the sensitivity distribution across the photodetector's focal surface.

Limiting ourselves to a flat-atmosphere model for the sake of simplifying the notation (generalization to a spherical model is not fundamentally difficult), under these assumptions the signal in an individual detector channel can be represented as an integral over the area of the elve observed by that channel
\begin{equation}
S_k[i]^{\mathrm{(m)}} \propto \frac{1}{h_\mathrm{e}^2} \int\int d\Omega\, dt\, \varepsilon_\mathrm{d}(\gamma, \psi)\, \left(\frac{dI}{dt}\right)^2 \, \sin^2\chi \cos^3\theta    
\end{equation}
where the integration is carried out over the spatiotemporal region bounded by the FoV of the $i$-th channel ($d\Omega = d\psi \sin\gamma d\gamma$) and the $k$-th detector time interval $T_k\leq T < T_{k+1}$ (the detector's spatiotemporal ``bin''). Here, $\chi$ is the angle between the dipole direction and the direction of the EMP from the source S$(x_0, y_0, z_0)$ to~E. Introducing the polar angles $\alpha_0$, $\phi_0$ and $\theta,\phi$ for these directions, respectively, we have the following.
\[
\cos\chi = \cos\theta \cos\alpha_0 + \sin\theta \sin\alpha_0 \cos(\phi-\phi_0).
\]
In the integrand, the substitution $\theta =  \theta(\gamma, \psi)$,  $\phi =  \phi(\gamma, \psi)$ must be performed.
\[
\tan\theta = \sqrt{h^2 \tan^2\gamma - 2h(x_0 \cos\psi - y_0 \sin\psi) + x_0^2 + y_0^2} / h_\mathrm{e}
\]
\[
\tan\phi = (h \tan\gamma \sin\psi - y_0) / (h \tan\gamma \cos\psi - x_0)
\]
where $h = H_\mathrm{d} - H_\mathrm{e}$ is the difference between the orbit altitude and the elve altitude (recall that $h_\mathrm{e} = H_\mathrm{e} - H_0$). Note that $t$ in~(3) is the time at the source. Due to the ``retardation effect'', the boundaries of the spatiotemporal bin become significantly nonlinear ($c$ is the speed of light).
\begin{equation}
t(T, \gamma, \psi) = T - T_0 - h_\mathrm{e}/(c\cos\theta) - h/(c\cos\gamma)
\end{equation}
The model signal (3) is necessary to form the likelihood function. To make the reconstruction more robust to potential inaccuracies in our model, following~\cite{Sharakin2024}, we will use Student's t-distribution for the likelihood function (with independently distributed signals for different times and different channels):
\begin{equation}
p(\cD_2|\Theta,\eta) = \prod_{i,k} \mathrm{StudentT}( S_k[i] | \mu = S_k[i]^{\mathrm{(m)}}, \sigma=\sigma_\mathrm{mes}, \nu = \nu_\mathrm{mes})
\end{equation}
where $\Theta = \{ x_0, y_0, z_0, \alpha_0, \phi_0, \tau_\mathrm{r}, \tau_\mathrm{d}, H_\mathrm{e} \}$, $\eta = \{ T_0, \sigma_\mathrm{mes}, \nu_\mathrm{mes}, \sigma_\mathrm{psf}, A_0 \}$.
Here, in addition to the measurement error model parameters $\sigma_\mathrm{mes}$ (characteristic error scale) and $\nu_\mathrm{mes}$ (parameter of ``non-Gaussianity'', where $\nu_\mathrm{mes} = 1$ corresponds to the Cauchy distribution), the list of auxiliary parameters includes $\sigma_\mathrm{psf}$, which allows for calculating the signal distribution across individual photodetector channels, and $A_0$, which controls the signal amplitude scaling factor ($A_0\propto I_{\max}^2$).
We have also included the altitude of the elve~$H_\mathrm{e}$ in the list of parameters of interest: the large volume of data per event (for a typical Mini-EUSO elve, the number of active channels can reach 400-600 or more, and the number of time steps~50-100) allows us to estimate~$H_\mathrm{e}$ alongside the discharge parameters.
In the model signal (3), the quantities~$h_\mathrm{e}$ and~$h$ now enter not only through the time delay~(4), as in the kinematic model, but also via the angles~$\theta$ and~$\chi$: the azimuthal distribution of the glow intensity along the ring is informative regarding both~$H_0$ and~$H_\mathrm{e}$ (and not just their combination~$h_\mathrm{e}$).

Unfortunately, the nonlinearity of the boundary of the spatiotemporal bin over which integration is performed in~(3) does not allow for a simple analytical approximation of the model signal. During the development of the elve, both the shape of this bin and its effective size change: numerical estimates show that at the beginning of the elve's development, the displacement of the ring boundary during the signal integration time exceeds the size of the channel's FoV, while at later stages it becomes significantly smaller.
A simple approximation of the integral in~(3) as the product of the integrand's value at the bin center and a (fixed) effective integration measure size (e.g., the product of the channel's FoV and the time step duration) is too crude and leads to significant biases in the posterior distribution—a good example of how substantial model inaccuracy can cause systematic reconstruction errors.

A more accurate evaluation of the integral in~(3) is possible using numerical methods, but implementing such an approach within a Bayesian model using probabilistic programming tools (e.g., in~PyMC-5~\cite{Abril-Pla2023}) is extremely challenging, and MCMC sampling in that case becomes very slow.

The problem of solving inverse problems through probabilistic inference without an explicit likelihood function – known as Likelihood-Free Inference~(LFI) – has become an active area of research in recent years. This field is now more commonly referred to as Simulation-Based Inference (SBI)~\cite{Cranmer2020}. The reason for this is clear: in many modern scientific and applied problems, it is difficult to formulate an explicit probability distribution linking model parameters and observations, $p(\cD|\Theta)$. However, it is possible to develop a procedure (a simulator) that, for a given model, generates a data sample for each parameter sample: $\cD_n = \mathrm{Simulator}(\theta_n)$.

Despite the absence of an analytically expressed likelihood function~$p(\cD|\Theta)$, the simulator enables probabilistic (Bayesian) inference by approximating, in one way or another, either the likelihood function~$p(\cD_\mathrm{obs}|\Theta)$ or directly the posterior distribution~$p(\Theta|\cD_\mathrm{obs})$.

SBI has been successfully applied in astronomy and cosmology (weak lensing, gravitational waves, galaxy cluster mass estimates, galaxy morphology determination, exoplanet atmosphere studies, etc., see references in~\cite{Ho2024}) and in particle physics~\cite{Brehmer2020}.

One of the most common SBI methods is ABC (Approximate Bayesian Computation)~\cite{Marin2012}, \cite{Sisson2018}. In ABC, samples from the posterior distribution are obtained as follows: first, a parameter is sampled from a proposal distribution (not necessarily identical to the prior). The data for this value are then generated (using a simulator). This data (or its summary statistics) is compared to the observed data using a certain distance metric. The acceptance or rejection of the proposed parameter value is determined by a pre-selected threshold.
The PyMC-5 library we use implements an ABC algorithm based on Sequential Monte Carlo (SMC), where the proposal distribution is sequentially refined. This significantly reduces the fraction of rejected samples but requires fast data simulators.

At the core of our stochastic generator lies Monte Carlo simulation: from the source point~S, electromagnetic wave ``beams'' $(\theta_n,\phi_n)$ are generated according to the directivity pattern of dipole radiation, with weights wn proportional to $|dI/dt|^2$. Subsequently, for each beam, the point~E of its intersection with the glow level – a sphere of radius $r_\mathrm{e} = R_\mathrm{E} + H_\mathrm{e}$ – is determined, and the arrival time~$T_n$ and direction $(\gamma_n, \psi_n)$ at the detector are calculated. The weighting coefficients are modified according to $w_n\to w_n \cos\gamma_n/l^2$, where $l = [1+h\tan^2\gamma_n/(2R_\mathrm{E})] h/\cos\gamma_n$ is the distance from the detector to the point~E.

To account for image blurring in the detector's optics, the direction of the incoming beam is randomly shifted according to a chosen Point Spread Function (PSF) (in the current implementation – an isotropic Gaussian distribution). Finally, based on $(T_n,\gamma_n,\psi_n)$, and in accordance with the FoV of the channels and the detector time discretization $T_k = k\tau$ (where $\tau$ is the signal integration time; in the case of the Mini-EUSO instrument, $\tau = 2.5$~$\mu$s), the corresponding spatio-temporal bin of the histogram~$S_k[i]$ is identified and the value~$w_n$ is added to~it.

This approach enables the calculation of data~$\cD_2$ for photodetectors with very different configurations, while accounting for both the non-uniformity of sensitivity distribution and the presence of structural ``dead'' zones between channels. It also facilitates straightforward modeling of signals with arbitrarily specified PSF, including tabulated PSF based on calibration experiments.

It is important to note that such a generator can be used both within SBI reconstruction (simulator mode) and for creating a database of model events (generator mode) for testing the methodology, including for controlling various model inaccuracies.
In both cases, to obtain statistically significant results when simulating the signal across all channels of the photodetector, the number of Monte Carlo beams can be very large (in the current version, $\sim10^8-10^9$ beams are used by default for the generator and $\sim10^6-10^7$ for the simulator). To accelerate the modeling process, the procedure described above has been implemented on a GPU using CUDA.

The stochastic generator allows for a detailed study of elve characteristics using model events. As mentioned earlier, a kind of information decoder is the projection of the data~$\cD_2$ onto the sphere of radius~$r_\mathrm{e}$ (the ``elve sphere''). Projection essentially involves mapping the measured signal values (taking into account the geometric factor $\cos\gamma/l^2$) ``backwards in time'', as it modifies the detector time~$T$ as\footnote{Here and henceforth, all ``decoded'' quantities, i.e., those obtained after inverse projection onto the elve sphere, are marked with the subscript~`e`.} $T_\mathrm{e}(T, \gamma) = T - l/c$ (cf.~(4)).

\begin{figure*}
    \centering
    \includegraphics[width=0.45\textwidth]{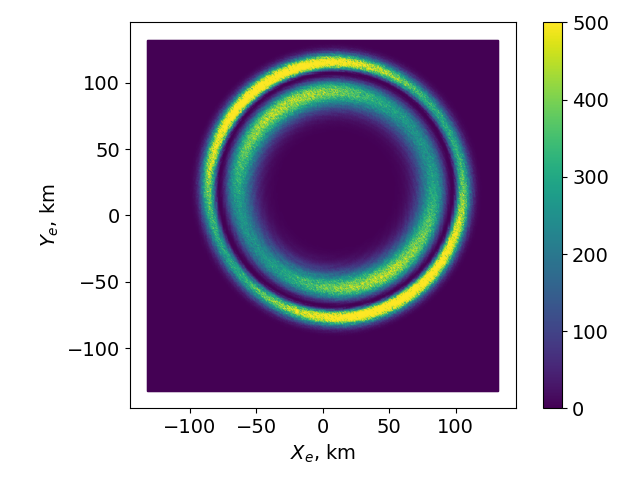}
    \includegraphics[width=0.45\textwidth]{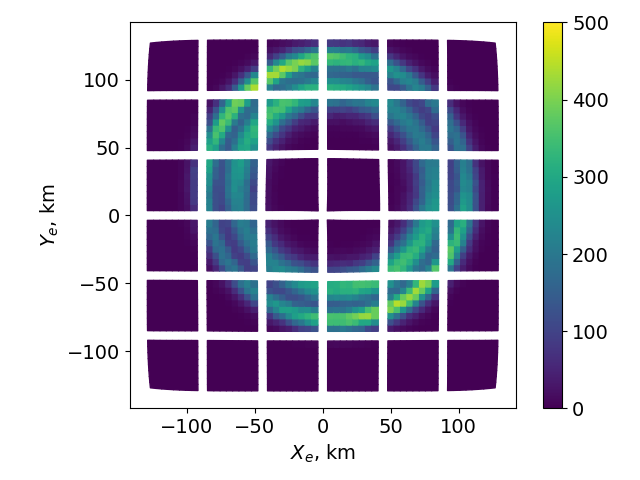}
    \caption{Projection of horizontal discharge with bigaussian current function ($\tau_\mathrm{r}=8\tau$, $\tau_\mathrm{d}=12\tau$). Left – for ideal detector, right – for Mini-EUSO.}
    \label{fig:Horiz_Frame}
\end{figure*}

It is precisely in this projection that the concentric nature of the expanding rings and the characteristic radial and azimuthal distributions are observed. Figure~\ref{fig:Horiz_Frame} shows one of the frames of a model event after projection onto the elve sphere, on the left for an ideal detector (with a very small pixel size and no optical aberrations), on the right for the Mini-EUSO detector (angular pixel size approximately~$0.8^\circ$, PSF size~$1^\circ$). The bigaussian discharge, with $\tau_\mathrm{r}=8\tau=20$~$\mu$s, $\tau_\mathrm{d}=12\tau=30$~$\mu$s, was oriented horizontally, $\alpha_0=90^\circ$, $\phi_0=30^\circ$, and located near the center of the FoV at $(x_0, y_0) = (10, 20)$~km at an altitude of~$H_0=5$~km. For ease of comparison, the signal is expressed as the surface density of the projected signal (in arbitrary units), which, within the framework of our adopted model of a proportional ionospheric response, corresponds to the electromagnetic energy flux density. The background fluctuations (noise) are assumed to be zero.

In both panels of the figure, the doublet structure of the elve is clearly discernible, associated with the presence of two fronts in the current function (the interval between the electromagnetic pulses is $\tau_\mathrm{r}+\tau_\mathrm{d}=20\tau=50$~$\mu$s). Precisely this scenario for the emergence of numerous elve doublets was initially proposed in~\cite{Marshall2012}. The reduction in energy density amplitude in the right panel is due to image blurring in the non-ideal optics of~Mini-EUSO.

Despite the image blurring (and its pixelation), this projection also reliably captures the azimuthal asymmetry associated with the orientation of the horizontal discharge: at $\phi_0=30^\circ$, the luminosity maxima of both rings lie in the regions $\Phi_\mathrm{e} = 120^\circ$ and~$300^\circ$.

\begin{figure*}
    \centering
    \includegraphics[width=0.45\textwidth]{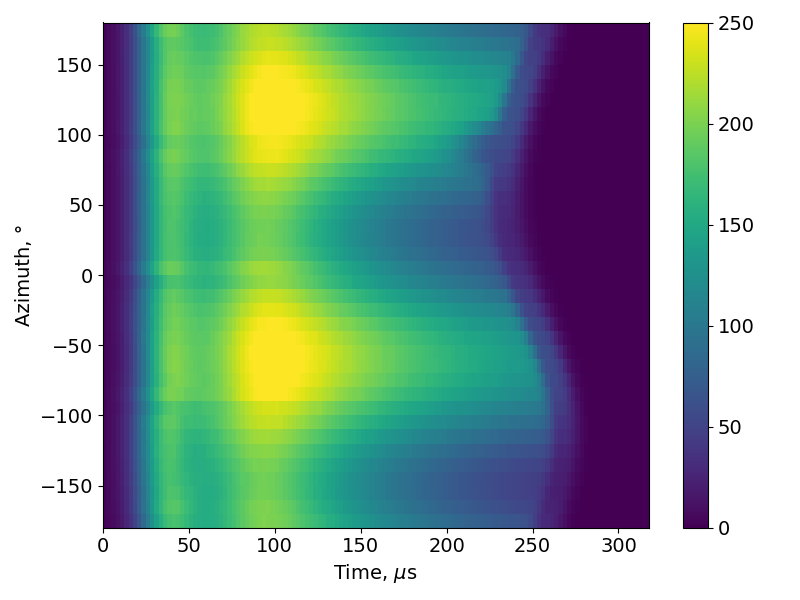}
    \includegraphics[width=0.45\textwidth]{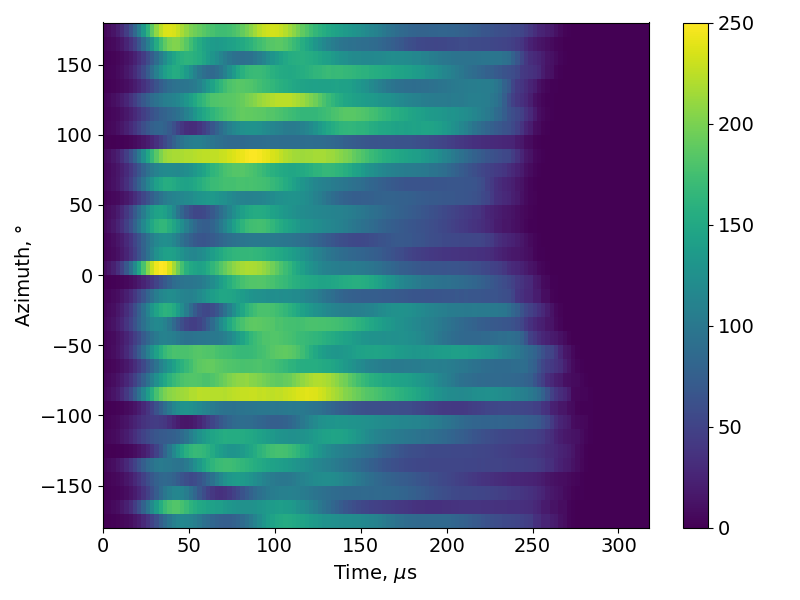}
    \caption{$\Phi_\mathrm{e} - T_\mathrm{e}$ histogram. Left – for ideal detector, right – for Mini-EUSO (Same event as one shown in previous figure).}
    \label{fig:PhiT}
\end{figure*}

To decode information about discharge orientation, it is convenient to represent the data as a two-dimensional histogram $\Phi_\mathrm{e}-T_\mathrm{e}$, as shown in Fig.~\ref{fig:PhiT}. To construct it, the center of the ring structure, the elve center E$_0(x_0, y_0)$, must first be determined. After that, the azimuthal angle values $\Phi_\mathrm{e}$ (with its vertex at~E$_0$) are divided into bins ($10^\circ$ in this case), which are then filled with the projected signal values. As in the previous figure, the histogram for an ideal detector is shown on the left, and the histogram for Mini-EUSO is shown on the right. The bin size for the time axis~$T_\mathrm{e}$ is chosen to be~$\tau$.

What stands out is the significant influence of the detector's instrument function on the pattern of the azimuthal histogram: the presence of dead zones in the instrument's FoV leads to a substantial transformation of the decoded pattern. This, in turn, complicates the reconstruction of the elve from such data using traditional optimization methods.

\begin{figure*}
    \centering
    \includegraphics[width=0.45\textwidth]{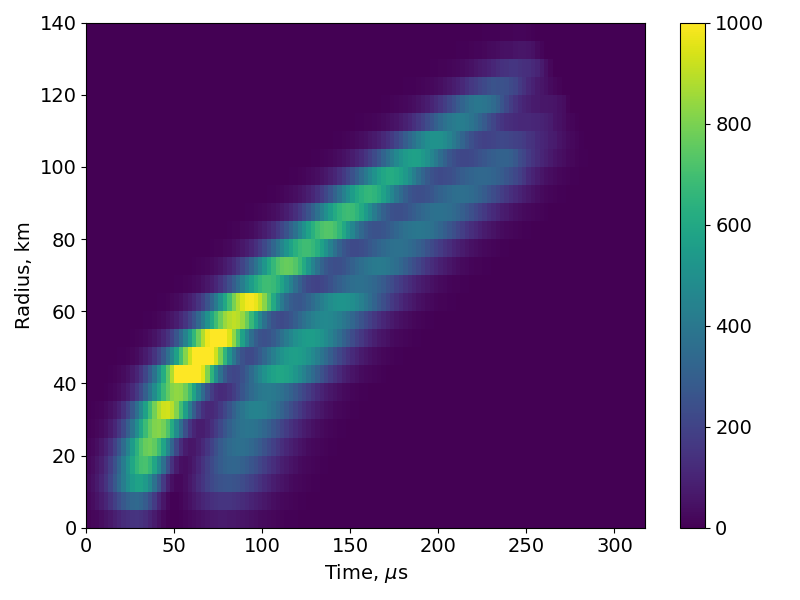}
    \includegraphics[width=0.45\textwidth]{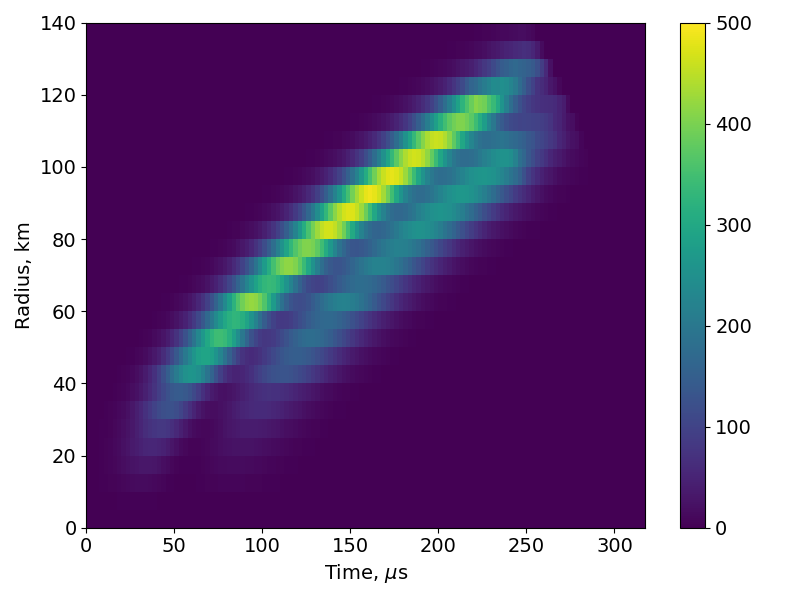}
    \caption{$R_\mathrm{e}-T_\mathrm{e}$ histogram. Left – for horizontally oriented discharge ($\alpha_0=90^\circ$), right – for vertical ($\alpha_0=0^\circ$).}
    \label{fig:RT}
\end{figure*}

A change in the discharge tilt angle~$\alpha_0$ leads to a shift in the signal's peak over time. This is clearly visible in Fig.~\ref{fig:RT}, which shows another type of polar histogram for the projected signal - the $R_\mathrm{e}-T_\mathrm{e}$ histogram. Here, $R_\mathrm{e}$ is the distance from the center of the projected FoV of the channel to the center of the elve, E$_0(x_0, y_0)$. The histogram on the left corresponds to a horizontal discharge ($\alpha_0=90^\circ$), while the one on the right corresponds to a vertical discharge ($\alpha_0=0^\circ$; all other discharge parameters are identical). The bin size is chosen according to the instrument's resolution: $\Delta R_\mathrm{e}=5$~km.

In both cases, for both rings of the doublet, the same relationship $R_\mathrm{e}^2 = c^2(T_\mathrm{e} - T_0)^2 - h_\mathrm{e}^2$  is evident (with different values of~$T_0$). However, the position of the maximum shifts to earlier moments~$T_\mathrm{e}$ (and smaller~$R_\mathrm{e}$) as $\alpha_0$ increases. We also draw attention to the difference in the color scale ranges: the elve luminosity is more intense for a horizontal discharge than for a vertical one (when a larger portion of the dipole's radiation energy does not reach the ionosphere).

Projecting a real measured signal onto the sphere $r = R_\mathrm{E} + H_\mathrm{e}$ introduces additional sources of uncertainty into the signal. These stem from both the presence of noise in the data and the process of histogramming (binning into spatio-temporal bins). However, all the noted regularities in the projected elve signal pattern are automatically accounted for within the Bayesian reconstruction framework based on the dynamic elve model. Furthermore, a subset of the model parameters is used to describe the additive noise, while another subset describes the instrument response function.

The validation of the dynamic elve reconstruction was performed using model events in a scenario of precise model specification. This means that the event was generated using the same stochastic generator that was subsequently used for its reconstruction, with broad, non-informative priors on all parameters.
For events similar to those recorded by the Mini-EUSO detector (with the same signal-to-noise ratio), and with a known~$H_\mathrm{e}$, the accuracy in recovering the discharge location is better than 1~km. The accuracy in recovering the dipole orientation strongly depends on the portion of the ring captured within the detector's FoV. To detect the azimuthal luminosity pattern, the algorithm requires data from different parts of the ring.
For the central events considered earlier, the posterior standard deviation for~$\alpha_0$ is about~$1^\circ$, and for~$\phi_0$, a few degrees. For peripheral events, where only a small segment of the elve ring falls within the FoV, the azimuthal angle cannot be reliably reconstructed. The reconstruction of the dipole tilt is more qualitative in nature in such cases, although distinguishing a horizontal from a vertical discharge does not pose significant difficulty.

\section{DISCUSSION}

For the reconstruction of discharge parameters from their ionospheric imprints (elves), proposed in this work, to be considered a fully-fledged method, further validation under conditions of various types of model misspecification is necessary. This refers, in particular, to the misspecification of the current function~$I(t)$.
This can be approached in two ways. On the one hand, one can formulate increasingly precise current model functions—with a larger number of parameters and greater physical justification. In discharge modeling, various modifications of the Transmission Line~(TL) model are often used (see~\cite{Watson2007}, \cite{Marshall2015}), which account for the wave-like nature of the propagation of the current pulse and the distribution of its amplitude along the line. For more details on common discharge current models, see~\cite{Karunarathne2014}.

On the other hand, if the goal of reconstruction is solely to estimate the dipole's location and orientation, it is advisable not to overly complicate the Bayesian model and to use only sufficiently ``coarse'' current profiles within it. In essence, the role of these current profiles here is the same as that of the model light curves discussed in our works~\cite{Sharakin2024} and~\cite{Sharakin2025} for meteor track reconstruction: through such imprecise (qualitative) profiling, we integrate the kinematic and dynamic (photometric) characteristics of the phenomenon into a single whole.
In situations where factoring different types of data is not advisable (for instance, due to ``data impoverishment'' at the edge of the FoV and within its dead zones), such a coarse approximation implements the necessary correlations into the Bayesian model. This allows for more efficient information extraction from the data and, consequently, leads to a more reliable (and accurate) reconstruction of the parameters of interest.

In the examples considered earlier, the doublet structure of the elve rings was caused by the presence of two fronts in the discharge current pulse. \cite{Marshall2015} presented an alternative viewpoint on the origin of doublets: the second ring arises from the interaction of the~EMP, reflected from the conductive Earth's surface, with the ionosphere. Such an interpretation is necessary when the measured temporal delay (i.e., the time interval after which the peak of the second ring passes through the same pixels as the peak of the first) significantly differs from the inter-pulse intervals typical in lightning discharges.

Furthermore, the magnitude of this time delay~$\Delta T$ allows an estimate of the discharge height~$H_0$. Indeed, in a flat atmosphere model:
\[
c\Delta T = \sqrt{R_\mathrm{e}^2+(H_\mathrm{e}+z_0)^2} - \sqrt{R_\mathrm{e}^2+(H_\mathrm{e}-z_0)^2}
\]
and under the approximation $H_0\ll H_\mathrm{e}$, we have:
$c\Delta T\approx 2z_0 /\sqrt{1+R_\mathrm{e}^2/H_\mathrm{e}^2}$
i.e., the distribution of delays is highly informative regarding the discharge height. This statement remains valid even when accounting for corrections due to atmospheric sphericity.

Large values of~$H_0$, for example $H_0 > 10$~km, can indicate that the cause of the doublet is a CID. In particular, most PIPER doublets were identified as reflected doublets generated by~CIDs. This could represent a significant advancement in the method, as it allows the separation of highly correlated parameters~$H_0$ and~$H_\mathrm{e}$ (in the absence of reflected signals in the data), allowing for independent estimation of each of these altitudes (rather than just their difference).

\begin{figure*}
    \centering
    \includegraphics[width=0.45\textwidth]{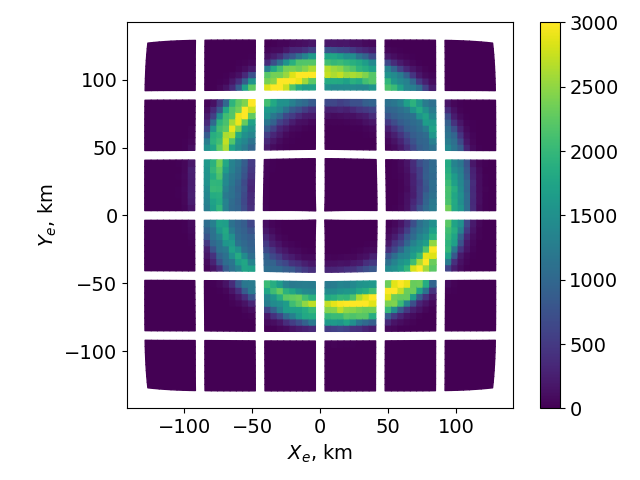}
    \includegraphics[width=0.45\textwidth]{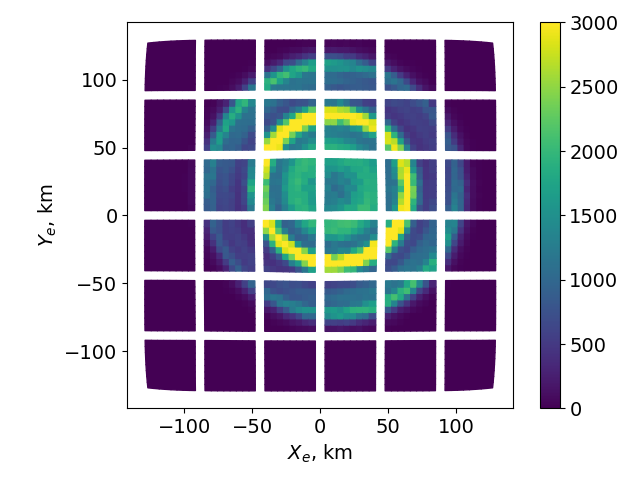}
    \caption{Comparison of image projections for two different heights of horizontal discharge: left – $H_0 = 5$~km, right – $H_0 = 15$~km, in a model accounting for reflection of~EMP (reflection coefficient equals to~1). Bi-Gaussian current function with $\tau_\mathrm{r} = 4\tau$, $\tau_\mathrm{d} = 8\tau$.
}
    \label{fig:REFL_Z05_Z15}
\end{figure*}

In Fig.~\ref{fig:REFL_Z05_Z15}, we present a comparison of model event images generated for two different $H_0$ values: 5~km (left) and 15~km (right; in both cases, $\alpha_0=90^\circ,\phi_0=30^\circ$). These illustrate how elves might appear for a typical intracloud discharge versus a CID. For simplicity, the electromagnetic pulse reflection coefficient from the Earth's surface is assumed to be~1, and a narrow bigaussian current profile is chosen ($\tau_\mathrm{r}=4\tau=10$~$\mu$s, $\tau_\mathrm{d}=8\tau=20$~$\mu$s).

For a low altitude discharge, all four rings (two each from the leading and trailing edges of direct and reflected electromagnetic pulses) merge into a single broad ring with a radius~$R_\mathrm{e}\sim90$~km. Its width is determined by the temporal parameters of the discharge and $H_0$ (as well as the size of the PSF). The time delay  $\Delta T \sim 24$~$\mu$s is slightly less than the interval between peaks in the current function, $\tau_\mathrm{r}+\tau_\mathrm{d}=30$~$\mu$s. At this stage of elve development, over such time intervals, the rings shift by only 2--2.5 pixels, which is heavily blurred by~PSF ($\sigma_\mathrm{psf}=1^\circ\sim 1.2$ pixels) and signal integration ($c\tau = 0.75$~km, ring displacement $\sim1$~km).

A different pattern is characteristic of an elve from a high-altitude discharge. At the time moment in the figure (approximately 150~$\mu$s after the first electromagnetic pulse peak), two rings from the current's leading edge are clearly identifiable. The time delay $\sim70-80$~$\mu$s is already sufficient for reliable separation of the rings in the image.

\begin{figure}
    \centering
    \includegraphics[width=0.45\textwidth]{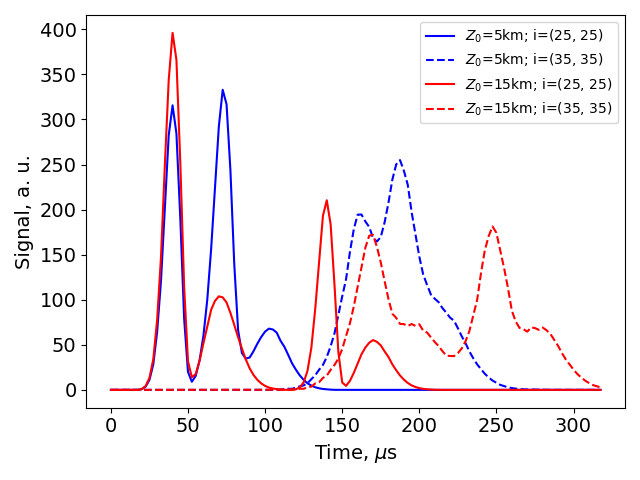}
    \caption{Comparison of signals in two detector channels, $i = (25, 25)$ – solid line and $(35, 35)$ – dashed line. Blue – for an event with $H_0 = 5$~km, red – with~$H_0 = 15$~km. (Dashed lines are presented with a scale factor of~5.)
}
    \label{fig:REFL_Z05_Z15_Signals}
\end{figure}

In reality, a varying number of rings can be identified in the image at different stages of elve development. For example, an event on Fig.~\ref{fig:REFL_Z05_Z15_Signals} with~$H_0=5$~km shows three rings in the early stages, while an event with~$H_0=15$~km shows four (the channel $i = (25,25)$\footnote{Here, the channel identifier~$i$ is indexed by the column number and the row number, with the origin $(0,0)$ located in the bottom-left corner of the projection map.} is located near the center of the elve). The ``disappearance'' of one ring in the first case is due to a specific parameter choice: the peak from the trailing edge of the direct electromagnetic pulse is close in time to the peak from the leading edge of the reflected pulse. This leads, among other things, to the second luminosity ring being more intense than the first—an effect repeatedly observed in Mini-EUSO data (despite using a reflection coefficient of~1, the added intensity is not very large because the amplitude ratio of the leading to trailing edge peaks in our model case is $\sim(\tau_\mathrm{r}/\tau_\mathrm{d})^2 = 4$).

In the same figure, the signal plots for channel $i=(35,35)$ correspond to a later stage of development (for clarity, these signals are scaled by a factor of~5 on the graph).

Thus, discharge altitude~$H_0$ significantly influences the luminosity pattern recorded by an imaging detector. Our proposed Bayesian model automatically tracks this influence in a quantitative manner and allows for (probabilistic) estimation not only of~$H_0$ but also of~$H_\mathrm{e}$.
As mentioned earlier, the simulator implements a simplified model of the electromagnetic pulse interaction with the ionosphere. In more advanced versions, this description could include more detailed characteristics of the ionospheric layer at altitudes of 80-95~km (for example, the altitude profile of electron concentration). This means that our proposed method of elve localization can be used not only to recover parameters of lightning discharges but also to conduct studies of the D-region ionosphere. The need to develop new methods for studying the D-region ionosphere, due to its insufficient understanding (despite a large number of experimental and theoretical works), is discussed, in particular, in~\cite{Kozlov2022}.

However, as the interpretational model becomes more complex, calling the simulator can become quite expensive, making reconstruction based on ABC inefficient. Alongside ABC, other SBI methods are increasingly being used in modern scientific research. These methods employ a neural network to estimate either the likelihood function (NLE, Neural Likelihood Estimation) or directly the posterior distribution density (NPE, Neural Posterior Estimation)~\cite{Lueckmann2021}.
Such SBI algorithms are amortized~\cite{Cranmer2020}, meaning they allow the application of a neural network pre-trained with the simulator to different events, thereby significantly reducing the time cost of each reconstruction. Furthermore, the numerous parameters of the neural network offer greater flexibility to improve the approximation accuracy compared to~ABC.

\section{CONCLUSION}

Lightning discharges, in addition to generating radio signals that propagate over long distances, leave their optical ``imprints'' in the ionosphere – elves. The dynamics of an elve is directly linked to the discharge parameters, primarily its location and orientation. An orbital imaging detector with wide-angle optics essentially functions as a large array of measurement stations (from several hundred to several thousand), collectively forming the spatiotemporal pattern of an event.
High-precision reconstruction of discharge parameters (especially for intracloud and CID discharges) and ionospheric parameters from these data is possible using Bayesian methods. The greatest potential in this direction, which we refer to as elves localization (``elveslocation''), lies with SBI – Simulation-Based Inference.

In~\cite{Iudin2018}, it was emphasized that satellite radio observations play a crucial role in studying processes in the thunderstorm atmosphere, providing a unique source of information on the spatiotemporal structure and directivity pattern of high-frequency radiation from various types of lightning discharge. The present work demonstrates that the analysis of ionospheric ``optical traces'' of discharges, recorded by multi-channel orbital detectors with microsecond resolution, can also make a significant contribution to building discharge models and refining the mechanism of lightning initiation.

\section{FUNDING}

The research was conducted under the state assignment of Lomonosov Moscow State University.



\end{document}